\documentclass[twocolumn,showpacs,amsmath,amssymb,aps,pra,floatfix]{revtex4}
\usepackage{bm}
\usepackage{amsfonts}
\usepackage{graphicx}
\begin{document}

\title{Relative variables of the exact fermion-antifermion bound state wave-function in the Schwinger Model}
\author{Tomasz Rado\.zycki}
\email{t.radozycki@uksw.edu.pl} \affiliation{Faculty of Mathematics and Natural Sciences, College of Sciences,
Cardinal Stefan Wyszy\'nski University, W\'oycickiego 1/3, 01-938 Warsaw, Poland}

\begin{abstract}
The exact Bethe-Salpeter amplitude for the fermion-antifermion bound state in the Schwinger Model is investigated. The dependence on the relative time and position in the center-of-mass frame in all contributing instanton sectors is analyzed. The same is accomplished for the relative energy and momentum variables. Several interesting properties of the amplitude are revealed. The explicit threshold structure is demonstrated.

\end{abstract}
\maketitle

\section{Introduction}

The formation of bound states is one of the most fundamental unsolved problems in quantum field theory. 
It becomes especially troublesome if relativistic nature of these states cannot be neglected and retardation effects in the interactions among constituent particles have to be taken into account. 

In the fifties of the last century, the quantum field theoretical equation for the bound state amplitude -- the so called Bethe-Salpeter (B-S)  equation -- was formulated~\cite{bs,gml} and the framework for considering bound states was established. However, for serious reasons, this equation cannot be solved in realistic quantum field theory. From the mathematical point of view, it is an extremely complicated multidimensional integral equation for a kind of a wave-function but the main difficulty comes from the fact, that from the very beginning one needs to know the nonperturbative propagators for ingredient particles and their interaction kernel. Obviously none of these quantities is known in theories like QED or QCD or even in simpler field theoretical models. 

The significant feature of the relativistic B-S amplitude is its dependence on the so called relative time, or relative energy, the role of which in characterizing a bound state has not yet become completely clear. Their presence in the theory mirrors the effects of retardation in the interactions and constitutes the main difference with respect to the nonrelativistic studies. It is also  a source of mathematical (and interpretational) complications.  

For the above reasons over the last sixty years much work has been devoted to the investigation of approximated versions of the equation~\cite{nn0,hl}. These approximations consisted on simplifying the propagators and interaction kernel (the most often the `ladder' approximation was used) and of reducing the equation to lower dimensions by neglecting the relative time or relative energy dependence~\cite{fg,bb,lusg,lus,bij,lucha}. In this case retardation effects and consequently a part of physical information were lost not allowing for the full study of the bound state properties as for instance its threshold structure. Some insight into the relative time (or energy) dependence of the bound system has been obtained in certain models~\cite{keister,ct,kopa,pd}. The wave-function dependence of relative variables was mainly given in euclidean space~\cite{wick,zuil,guth,efi,dork2,dork1,carb1}, where it is easier to obtain, but the Minkowski-space approach for a bound system has also been developed~\cite{kusaka, kusaka1,kara,kara1,carb1,carb2,carb3,carb4,hall}.

A very unpleasant situation is that full results are still lacking even in simple field theoretical models. There are only few examples, as for instance the so-called Wick-Cutkosky Model~\cite{wick,cut,nn1}, where it was possible to find `exact' solutions (but with the `ladder' approximation embodied into the model). In this context we would like to mention the result obtained in the Schwinger Model (SM) -- a two-dimensional massless electrodynamics~\cite{js} -- where {\em full} (i.e. {\em exact}) B-S wave-function was found, in the spirit of~\cite{eden,mand}, from the residue of the pole corresponding to the bound particle, but also as a direct computation of the field amplitude~\cite{trsing,trfactor} opening thereby the possibility of its investigation. 

To our knowledge, no other truly exact solutions of the B-S equation in nontrivial field theoretical models are known. Our main motivation in this work is then the examination of the previously found B-S amplitude from the point of view of the relative time and energy dependence. One should stress here that the considered model is fully relativistic, since it deals with massless fermions. 

Although B-S function does not have any direct probabilistic interpretation, it appears in the matrix elements and scattering processes involving bound states~\cite{mand, hua}. The results may be then of certain interest both in particle and nuclear physics. As pointed out in~\cite{keister} the relative time or energy dependence is important for the processes in which the bound state appears in loop integrations, but also can manifest itself when one deals with high momentum transfers. The results, which in the SM are exact, may constitute also the measure of accuracy of various approximation schemes.

The Schwinger Model offers one more attractive possibility. As it is well known, it is the theory with nontrivial vacuum~\cite{cadam1,smil,gattr,maie,rot,gmc}, which has the character of the $\theta$-vacuum. Apart from the confinement of fermions this feature resulting from the the existence of instanton sectors makes the theory to be a simple model of QCD. The nonzero instanton sectors do contribute to the B-S amplitude, and the appropriate contribution was explicitly found in~\cite{trsing}. One can then analyze how instantons influence the relative energy and momentum dependence of the bound state wave-function (an attempt to account for instanton effects in bound states was made in~\cite{car,duncan}).

The present paper is organized as follows. In the next section we invoke the exact B-S amplitude in coordinate space for all contributing instanton sectors. We investigate its position dependence for fixed values of relative time and {\em vice versa} in the center-of-mass frame and present the appropriate plots. In section~\ref{sec:bsm} we concentrate on the behavior of the amplitude in the relative momentum for certain characteristic values of relative energy again in the center-of-mass frame and we depict the results of the numerical computations for each sector. At the end we present the density plots revealing the dependence of the wave-function on the whole two-momentum argument and show the trajectories of the threshold cusps. And finally in the last section we give the summary of the obtained results and propose some conclusions.

\section{Bethe-Salpeter amplitude in coordinate space}
\label{sec:bsc}

In order to fix the notation we start this section with recalling basic facts concerning SM. Its Lagrangian density is given by
\begin{eqnarray} 
{\cal L}(x)=&&\!\!\!\!\!\!\overline{\Psi}(x)\bigg(i\gamma^{\mu}\partial_{\mu}-
gA^{\mu}(x)\gamma_{\mu}\bigg)\Psi (x)\nonumber\\
&&\!\!\!\!\!\!- \frac{1}{4}F^{\mu\nu}(x)F_{\mu\nu}(x)- \frac{\lambda}{2}\left(\partial_{\mu}A^{\mu}(x)\right)^2, 
\label{eq:lagr} 
\end{eqnarray} 
where $g$ is the coupling constant and $\lambda$ the gauge fixing parameter (in what follows we choose Landau gauge setting
$\lambda\rightarrow\infty$). Dirac gamma matrices are chosen in two-dimensional form:
\begin{eqnarray}
&&\gamma^0=\left(\begin{array}{lr}0 & \hspace*{2ex}1 \\ 1 & 0 
\end{array}\right), \;\;\;\;\; 
\gamma^1=\left(\begin{array}{lr} 0 & -1 \\ 1 & 0 
\end{array}\right),\nonumber\\ 
&&\gamma^5=\gamma^0\gamma^1=\left( 
\begin{array}{lr} 1 & 0 \\ 0 & -1 \end{array}\right),
\label{eq:gammas}
\end{eqnarray}
and for the metric tensor we have $\mathfrak{g}^{00}=-\mathfrak{g}^{11}=1$. In the Schwinger's papers~\cite{js} the
nonperturbative formula for the fermion propagator in coordinate space was found. It may be given the following form:
\begin{equation}  
S(x)={\cal S}_0(x)\exp\left[-ig^2\beta(x)\right],  
\label{eq:propbet} 
\end{equation} 
with ${\cal S}_0(x)$ being the free propagator:
\begin{equation}
{\cal S}_0(x)=-\frac{1}{2\pi}\ \frac{\not\! x}{x^2-i\varepsilon}
\label{eq:s0}
\end{equation}
and function $\beta$ defined by  
\begin{eqnarray}  
&&\beta(x)=\label{eq:beta}\\
&&\left\{\begin{array}{l}\frac{i}{2g^2}\left[- 
\frac{i\pi}{2}+\gamma_E +\ln\sqrt{ \mu^2x^2/4}+   
\frac{i\pi}{2}H_0^{(1)}(\sqrt{\mu^2x^2})\right], \\
\hspace*{10ex}
x\;\;\;\; {\rm timelike},\\   
\frac{i}{2g^2}\left[\gamma_E+\ln\sqrt{- 
\mu^2x^2/4}+K_0(\sqrt{-\mu^2x^2})  
\right], \\ \hspace*{10ex} x\;\;\;\; {\rm  
spacelike}.\end{array}\right.  
\nonumber  
\end{eqnarray}  
Symbol $\gamma_E$ denotes here the Euler constant, $\mu$ is the Schwinger boson mass ($\mu^2=g^2/\pi$) and 
$H_0^{(1)}$ and $K_0$ are Hankel function of the first kind and Basset
function respectively.
 
In the further development of the model higher Green's functions in explicit form were obtained. Special attention was paid to two-fermion function~\cite{lsb,trjmn,trinst}. The analysis of its analytical structure showed that it  contains the bound state pole with the residue defining the B-S amplitude~\cite{trsing,trfactor}. It has the contributions from $k=0$ and $k=\pm 1$ instanton sectors and may be given the form:
\begin{equation}
\Phi_P(x)=\Phi_P^{(0)}(x)+\Phi_P^{(1)}(x),
\label{eq:phip}
\end{equation}
where 
\begin{eqnarray}
\Phi_P^{(0)}(x)&\!\!\! =&\!\!\! -2\sqrt{\pi}S(x)\gamma^5\sin(Px/2),\label{eq:phip0}\\
\Phi_P^{(1)}(x)&\!\!\! =&\!\!\! \frac{\mu}{2\sqrt{\pi}}e^{\gamma_E}e^{ig^2\beta(x)}e^{-i\theta\gamma^5}\gamma^5\cos(Px/2)\; .\label{eq:phip1}
\end{eqnarray}
$P$ denotes here the total two-momentum of the bound state (satisfying $P^2=\mu^2$), $x=[t,r]$ is the relative two-position variable and $\theta$ is the parameter defining the $\theta$-vacuum.

The bound state amplitude is then defined entirely by the function $\beta$ and trigonometric functions. The presence of these functions (sine for $k=0$, and cosine for $k=\pm 1$) shows that there are regions where noninstantonic sector dominates, and other, where instanton contributions are larger.  

\begin{figure*}
\centering
{\includegraphics[width=1\textwidth]{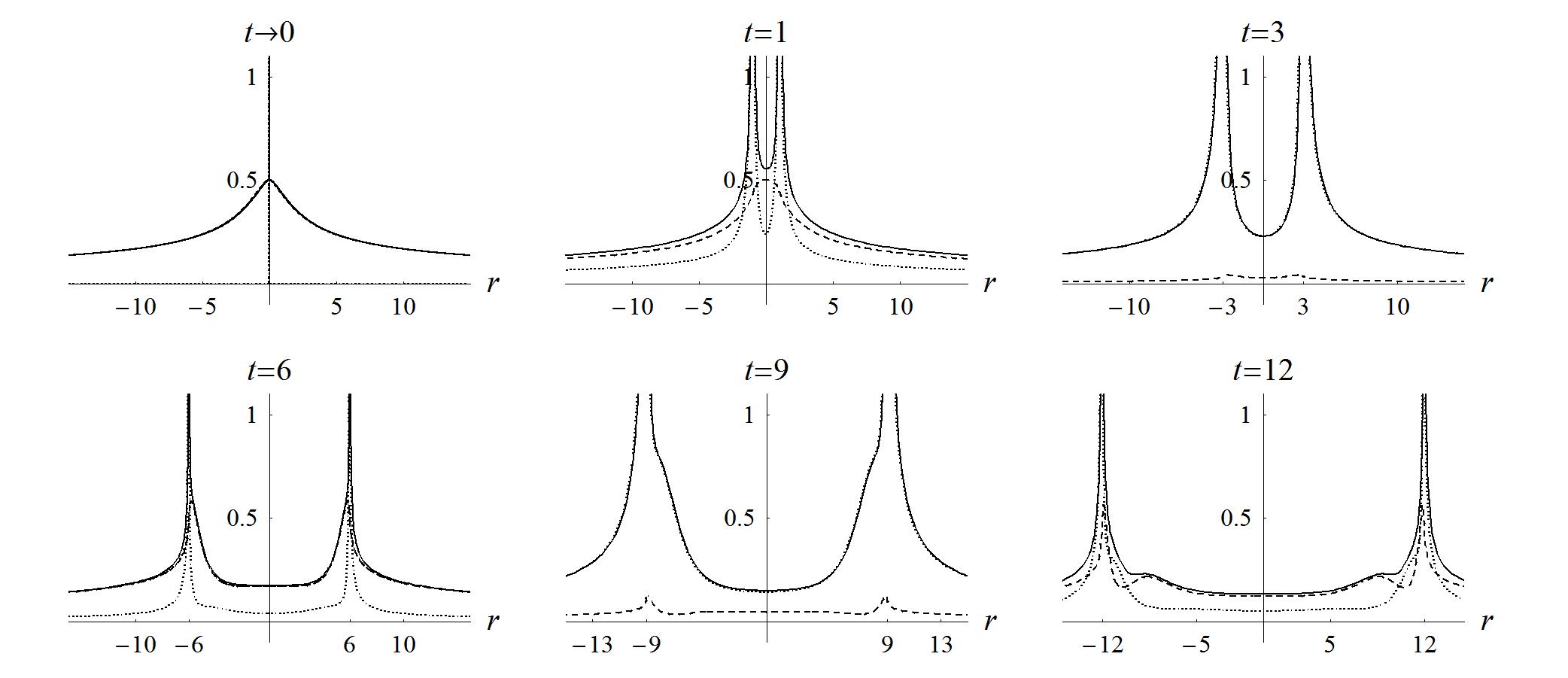}
\caption{The dependence of amplitudes $|\Phi_P^{(0)}|$ (dotted line), $|\Phi_P^{(1)}|$ (dashed line) and $|\Phi_P|$ (solid line) in units of $\mu$ on  relative position $r$ with fixed relative time $t$. The variable $t$ for the successive plots is chosen to be: $0$, $1$, $3$, $6$, $9$, $12$. All coordinates are in units of $\mu^{-1}$. The dotted line for $t=3$ and $t=9$ is covered by the solid line since the instanton contribution is tiny.} \label{fig:plx}}
\end{figure*}

For our further analysis we choose the center-of-mass frame in which $P=[\mu,0]$. In such a case the trigonometric functions depend only on the relative time and expressions~(\ref{eq:phip0}) and~(\ref{eq:phip1}) reveal quasi-periodic structure in this variable. For $t$ approaching $(2n+1)\pi/\mu$ the amplitude $\Phi_P^{(0)}$ dominates over topological part, but for $t$ close to $2n\pi/\mu$ the instanton corrections become more important (except the light cone).  

If we fix the relative time to take certain chosen value, the dependence of the amplitude on the spatial variable $r$ may be easily plotted. Since $\Phi_P^{(0)}$ and $\Phi_P^{(1)}$ are \mbox{$2\times 2$} complex matrices, we will draw the behavior of the quantities:
\begin{eqnarray}
|\Phi_P^{(0)}|&\!\!\! =&\!\!\! \left(\frac{1}{2}\ \mathrm{tr}[\Phi_P^{(0)+}\Phi_P^{(0)}]\right)^{1/2},\label{eq:modp0}\\
|\Phi_P^{(1)}|&\!\!\! =&\!\!\! \left(\frac{1}{2}\ \mathrm{tr}[\Phi_P^{(1)+}\Phi_P^{(1)}]\right)^{1/2},
\label{eq:modp1}
\end{eqnarray}
without any claim to their probabilistic interpretation. With this definition the dependence of $|\Phi_P^{(1)}|$ on the unphysical $\theta$-parameter (after all in the SM it may be gauged away) disappears. The real and imaginary parts of~(\ref{eq:phip0}) and~(\ref{eq:phip1}) may also be easily plotted if needed. Thanks to the trace properties of Dirac gamma matrices, for the total amplitude, we get
\begin{equation}
|\Phi_P|= \sqrt{\frac{1}{2}\ \mathrm{tr}[\Phi_P^{+}\Phi_P]}=\sqrt{|\Phi_P^{(0)}|^2+|\Phi_P^{(1)}|^2}.
\label{eq:topi}
\end{equation}

In Figure~\ref{fig:plx} the dependence of both contributions together with~(\ref{eq:topi}) on the relative position $r$ for fixed relative time is plotted. For $t\rightarrow 0$ the amplitude $|\Phi_P^{(0)}|$ (which is indicated in all graphs with a dotted line) is hardly visible, since it becomes a Dirac delta function, which is not graceful for drawing. Due to the denominator in~(\ref{eq:s0}) the amplitude for $k=0$ exhibits the singularities on the light cone. This does not refer to the nonzero instanton sector (dashed line). The solid line corresponds to the expression $|\Phi_P|$. 

On the third plot (i.e. for $t=3$ in units of $\mu^{-1}$) we observe the broadening of peaks in $|\Phi_P^{(0)}|$ and strong reduction of the value of $|\Phi_P^{(1)}|$. This is understandable since $3$ is close to $\pi$ (hence $\sin(\mu t/2)\approx 1$ and $\cos(\mu t/2)\approx 0$). The same refers to $t=9$, which approaches $3\pi$. On the other hand, for $t=6$ (i.e. close to $2\pi$) and $t=12$ (i.e. almost $4\pi$) $|\Phi_P^{(1)}|$ becomes more important, and $|\Phi_P^{(0)}|$ is close to zero, apart from the proximity of the light cone, where peaks turn now very narrow. This kind of light-cone singularities is connected with the masslessness of the fundamental fermions in the model.

\begin{figure*}
\centering
{\includegraphics[width=1\textwidth]{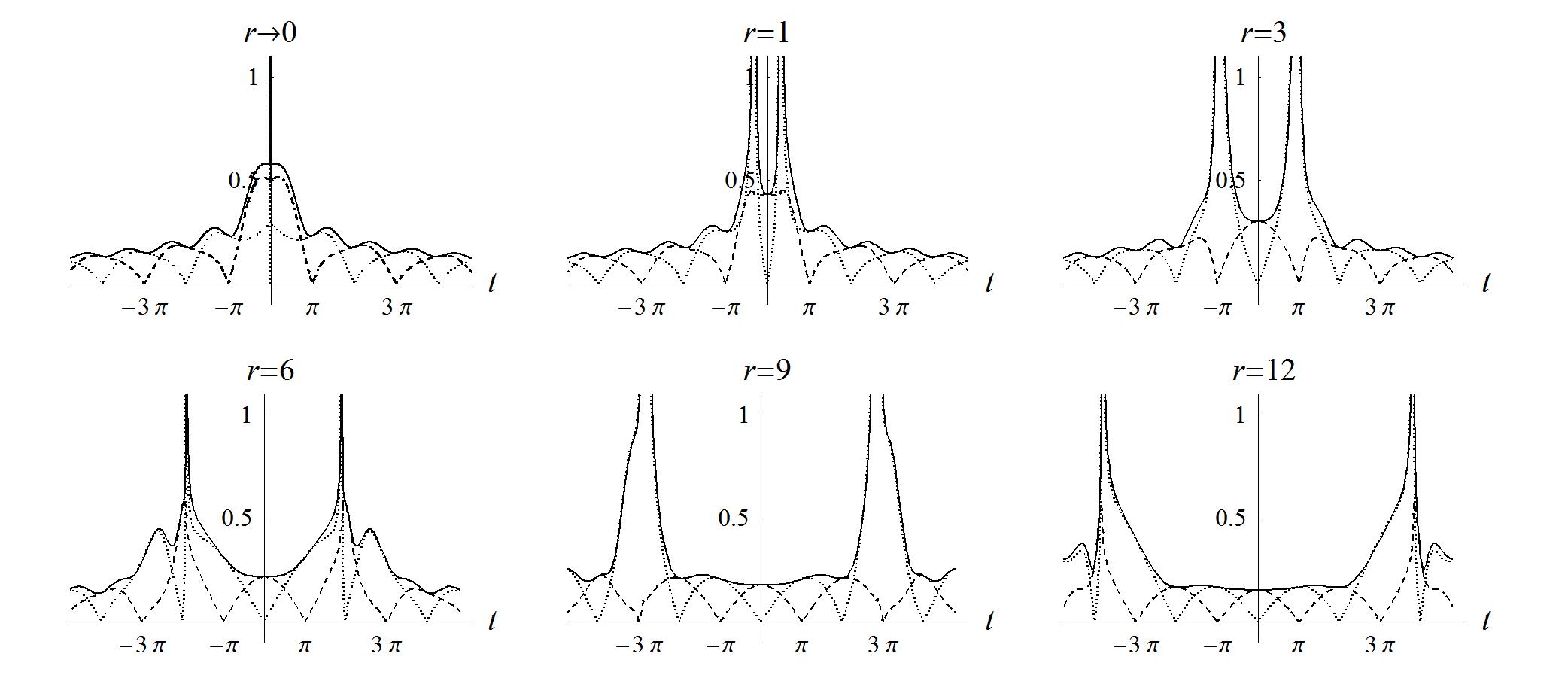}
\caption{Same as Figure~\ref{fig:plx}, but with fixed relative position $r$. The variable $r$ for the successive plots is chosen to be: $0$, $1$,  $3$, $6$, $9$, $12$.} \label{fig:plt}}
\end{figure*}

Since the Basset function tends to zero for large arguments:
$$K_0(z)\sim e^{-z}/\sqrt{z},$$
 it may be easily seen from~(\ref{eq:phip0}) and~(\ref{eq:phip1}) that both amplitudes behave like  $1/\sqrt{|r|}$ and become comparable to each other for large separations. This is a relatively slow decrease and is again related to the formation of the massive bound state by massless fermions and to the presence of fermionic zero modes in the instanton background. One should also note, that in that region (strictly speaking outside the light cone) the whole amplitude in both sectors is real (for $\theta=0$). 

If we now fix the relative position and investigate in an analogous way the relative time dependence of the amplitudes, we expect their quasi-periodic behavior due to to the presence of trigonometric functions (in a boosted frame, this character would appear also in the $r$ dependence). They make $|\Phi_P^{(0,1)}|$  to have (by turn) nodes at the integer values of $\pi/\mu$. This can be seen in Figure~\ref{fig:plt}, together with the typical light-cone singularities. Again the peaks of $|\Phi_P^{(0)}|$ widen for odd values of $\pi/\mu$ (what is also in general accompanied by the damping of instanton contributions) and become sharp for even ones. The latter are regions of instanton domination. On the other hand $|\Phi_P|$ is much more smooth with no nodes at all. For example if the relative distance gets large, there is for relatively small $t$ a whole range of almost constant $|\Phi_P|$. One can easily check, that in this case:
\begin{eqnarray}
|\Phi_P^{(0)}(x)|&\!\!\! =&\!\!\! \sqrt{\frac{\mu}{2\pi}} e^{\gamma_E/2} e^{\frac{1}{2} K_0(\sqrt{-\mu^2 x^2})}|\sin(Px/2)|,\label{eq:mopi}\\
|\Phi_P^{(1)}(x)|&\!\!\! =&\!\!\! \sqrt{\frac{\mu}{2\pi}} e^{\gamma_E/2} e^{-\frac{1}{2} K_0(\sqrt{-\mu^2 x^2})}|\cos(Px/2)|.
\nonumber
\end{eqnarray}
Since for those fairly large values of $-x^2$ the function $K_0(\sqrt{-\mu^2 x^2})$ equals almost zero, then both contributions complement each other and $|\Phi_P|$ becomes a very slowly varying function. This interplay between oscillating noninstantonic and instantonic parts constitutes an interesting observation.

For large values of relative time (i.e. for large timelike separations) we have to use the upper formula of~(\ref{eq:beta}). The asymptotic behavior is then dictated by the presence of a logarithm since for the Hankel function we have: 
$$H_0^{(1)}(z)\sim e^{iz}/\sqrt{z}\underset{z\rightarrow\infty}{\longrightarrow}0$$
and we conclude that the amplitudes fade out as $1/\sqrt{t}$, hence similarly to the previous case. 

\section{Bethe-Salpeter amplitude in momentum space}
\label{sec:bsm}

\subsection{$k=0$ instanton sector}
\label{subsec:k0}

In real four-dimensional quantum field theory calculations one mostly has to do with Feynman diagrams in momentum space. In this representation the $S$-matrix elements have their natural form and the analytical properties of the
Green's functions (poles, cuts) are closely related to the physical quantities. One can expect that also B-S wave-function should reveal certain important properties regarding for instance the internal threshold structure.

To analyze the dependence of B-S wave-function on the relative two-momentum variable $Q=[E,q]$, what we need, is the transformed propagator~(\ref{eq:propbet}), since
\begin{eqnarray}
\Phi_P^{(0)}(Q)&\!\!\! =&\!\!\! -2\sqrt{\pi}\int d^2x e^{iQx}\sin(Px/2)S(x)\gamma^5\label{eq:p0q}\\
&\!\!\! =&\!\!\! i\sqrt{\pi}\left[S\left(\frac{P}{2}+Q\right)-S\left(-\frac{P}{2}+Q\right)\right]\gamma^5.
\nonumber
\end{eqnarray}
Due to its complicated form it cannot be, however, explicitly Fourier-transformed. The direct numerical transformation is, on the other hand, very slowly convergent due to the oscillatory nature of integrand function. For these technical reasons, we prefer to make use of the integral representation of the euclidean version of propagator $S(p)$~\cite{trpert}, which may be in the straightforward way continued to spacelike momenta:
\begin{eqnarray}
S(p)&\!\!\!=&\!\!\!\sqrt{\frac{\mu}{2}}\frac{\not\! p}{(-p^2)^{5/4}}e^{\gamma_E/2}\label{eq:sspac}\\
&&\!\!\!\!\!\!\times\int_0^{\infty}dx \sqrt{x}J_1(x)\exp\left[\frac{1}{2}K_0\left(\mu  
x/\sqrt{-p^2}\right)\right],
\nonumber
\end{eqnarray}
where $J_1$ is the Bessel function.
One might say that in this formula the angular part of the two-dimensional transform has been performed, and the radial part is still left. To make the appropriate plots we will need to know the values of $S(p)$ in the whole momentum plane. The analytical continuation of~(\ref{eq:sspac}) to the region, where $p^2>0$, gives:
 \begin{eqnarray}
S(p)&\!\!\!=&\!\!\!-\frac{\sqrt{\mu}}{2}\frac{\not\! p}{(p^2)^{5/4}}(1-i)e^{\gamma_E/2}\label{eq:stim}\\
&&\!\!\!\!\!\!\times\int_0^{\infty}dx \sqrt{x}J_1(x)\exp\left[\frac{i\pi}{4}H_0^{(1)}\left(\mu  
x/\sqrt{p^2}\right)\right].
\nonumber
\end{eqnarray}
Using together~(\ref{eq:sspac}) and~(\ref{eq:stim}) we get the integral representation of~(\ref{eq:p0q}). 

\begin{figure}[h]
\centering
{\includegraphics[width=0.45\textwidth]{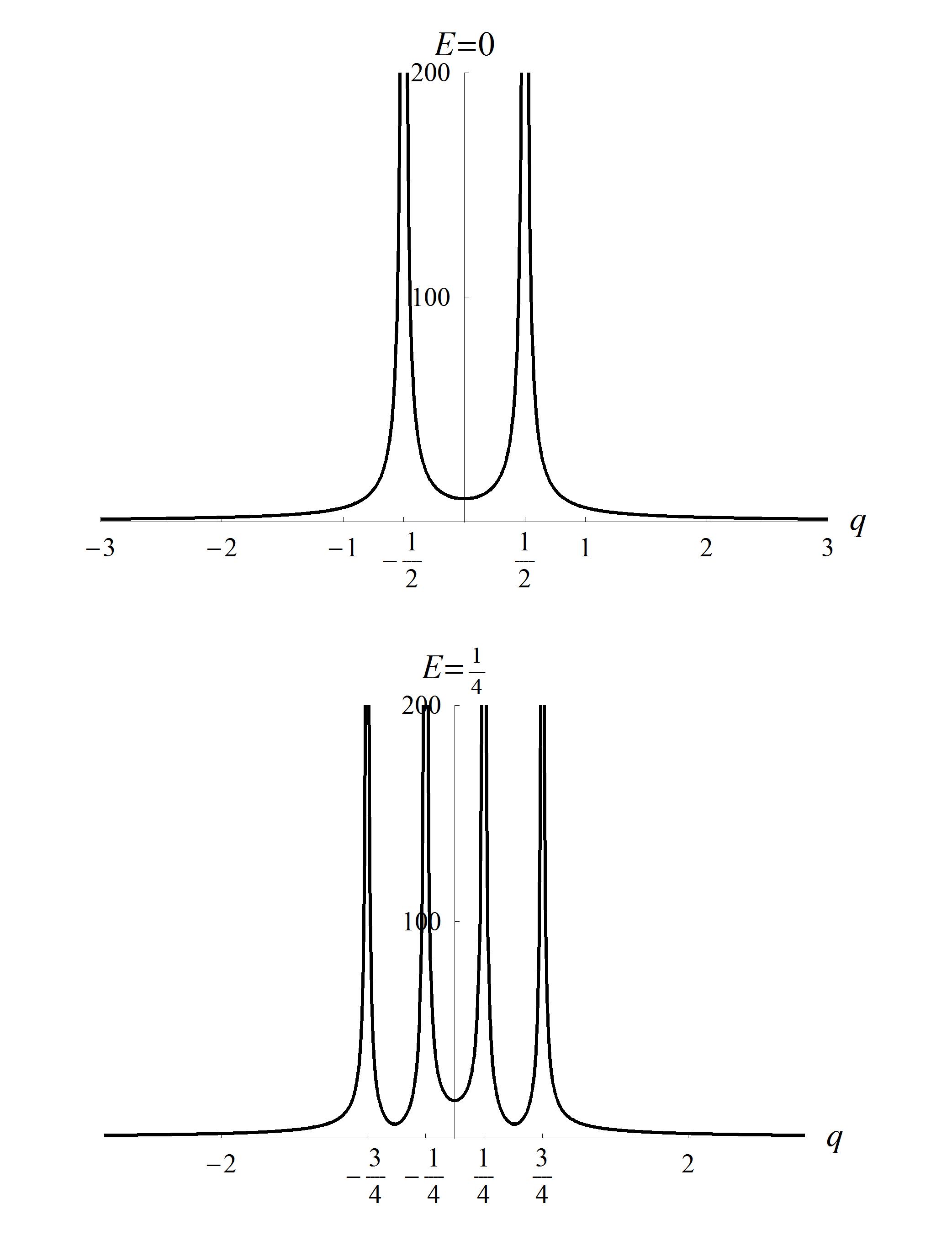}
\caption{The dependence of the amplitude $|\Phi_P^{(0)}|$ (in units of $\mu^{-1}$) on the relative momentum $q$ for fixed values of relative energy (in $\mu$) : $E=0$ (upper plot) and $E=1/4$ (lower plot).} 
\label{fig:plni12}}
\end{figure}

Figure~\ref{fig:plni12} shows the dependence of the quantity $|\Phi_P^{(0)}|$ defined again by the formula~(\ref{eq:modp0}), (but referring now to the object~(\ref{eq:p0q})) on the relative momentum for two chosen values of relative energy: $E=0$ and $E=1/4\  \mu$ in the center-of-mass frame.

The observed singularities are of kinematical nature (similar ones appear in~\cite{carb2,carb3,carb4}) and correspond to the external `legs' of the B-S wave-function ($\Phi_P^{(0)}$ is not an amputated function). They arise, when \mbox{$(P/2+Q)^2=0$} or $(-P/2+Q)^2=0$. These two conditions may be rewritten as
\begin{eqnarray}
&&(\mu/2+E)^2-q^2=0,\nonumber\\
&&(-\mu/2+E)^2-q^2=0.
\label{eq:con0}
\end{eqnarray}
Hence for $E=0$ the singularities are situated at $q=\pm 1/2\ \mu$. If relative energy increases, four such singularities appear. For instance at $E=1/4\ \mu$, we get $q=\pm 1/4\ \mu$ and $q=\pm 3/4\  \mu$. This is clearly visible on the second plot of Figure~\ref{fig:plni12}.

\begin{figure}[h]
\centering
{\includegraphics[width=0.45\textwidth]{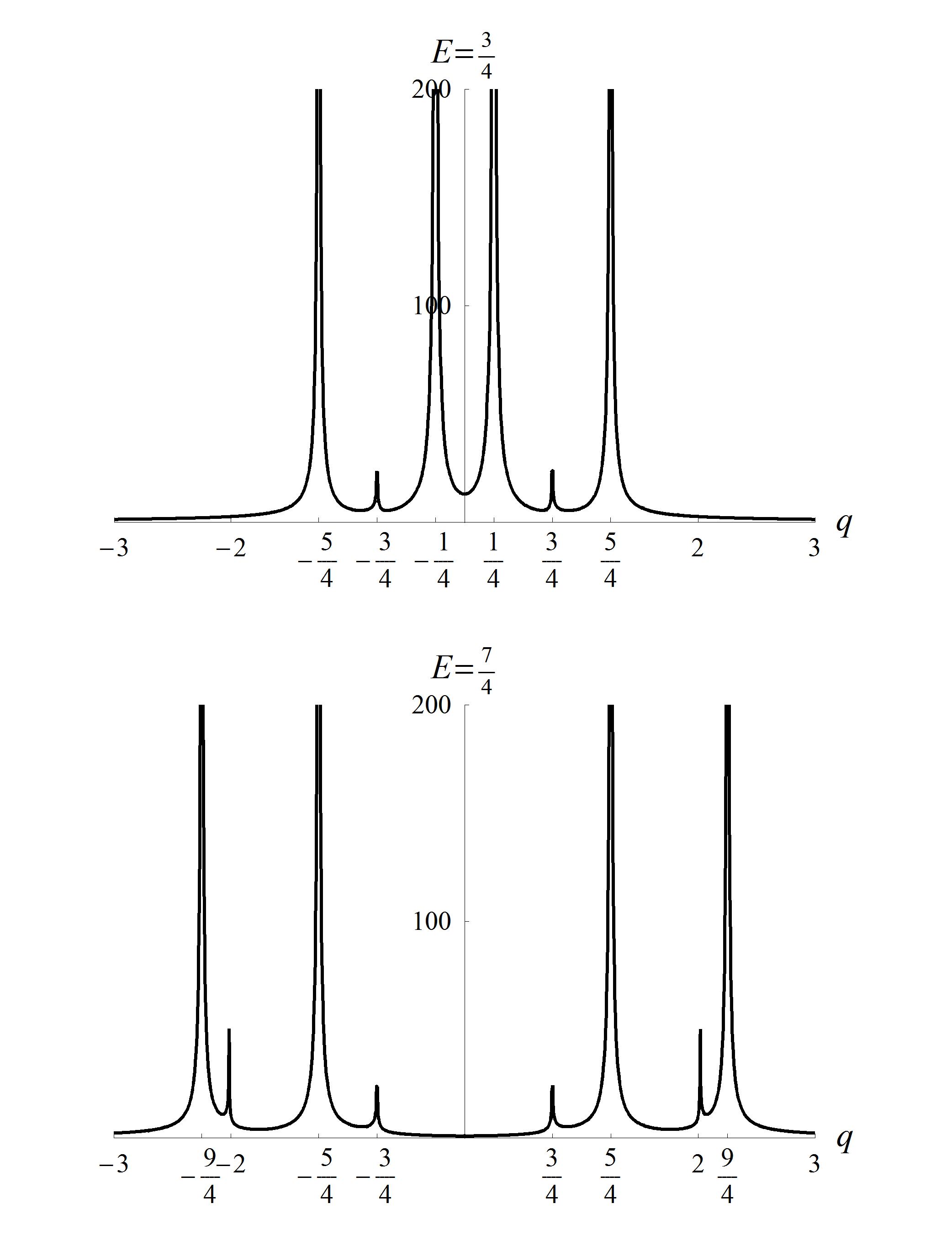}
\caption{Same as in Figure~\ref{fig:plni12} but for $E=3/4$ (upper plot) and $E=7/4$ (lower plot).} \label{fig:plni34}}
\end{figure}

When the relative energy reaches the threshold corresponding to the Schwinger boson, a new effect appears. In general $n$-th order thresholds result from satisfying the equations: 
\begin{eqnarray}
&&(P/2+Q)^2=(n\mu)^2\nonumber\\
&&(-P/2+Q)^2=(n\mu)^2,
\label{eq:ntre}
\end{eqnarray}
i.e. when appropriately high momentum transfer becomes possible. In the center-of-mass frame their appearance requires that $(\pm \mu/2+E)^2-n^2\mu^2>0$ and hence:
\begin{equation}
\left[E-\left(n\mp\frac{1}{2}\right)\mu\right]\left[E+\left(n\pm\frac{1}{2}\right)\mu\right]>0.
\label{eq:coe}
\end{equation}
For $n=1$ we see that, while taking upper signs in~(\ref{eq:coe}), the first two cusps should appear, if $E>1/2\  \mu$. For instance as $E=3/4\  \mu$, the solutions of the first equation of~(\ref{eq:ntre}) are 
\begin{equation}
q=\pm\sqrt{\left(E-\frac{1}{2}\ \mu\right)\left(E+\frac{3}{2}\ \mu\right)}=\pm \frac{3}{4}\  \mu.
\label{eq:q1}
\end{equation}
These cusps are clearly visible on the first plot of Figure~\ref{fig:plni34} in places exactly predicted, apart from kinematic singularities at $q=\pm 1/4\  \mu$ and $q=\pm 5/4\  \mu$ resulting from~(\ref{eq:con0}).

Moving further we see, that if relative energy exceeds the value of $3/2\ \mu$, the condition~(\ref{eq:coe}) with lower signs may be satisfied too. This should produce additional two cusps. For exemplary value $E=7/4\ \mu$ it may be easily found that the positions of all four cusps are
$$
q=\pm \frac{3}{4}\ \mu,\;\;\;\; q=\pm\frac{\sqrt{65}}{4}\ \mu\approx \pm 2.02\ \mu.
$$
This is exactly, what we see on the second plot of Figure~\ref{fig:plni34} (beside kinematic singularities which now moved to $\pm 5/4\ \mu$ and $\pm9/4\  \mu$).

Putting $n=2$ into~(\ref{eq:coe}) we find that second order cusps connected with two-boson thresholds appear for $E>3/2\ \mu$ (the first two) and for $E>5/2\ \mu$ (the other two). These cusps are however extremely weak (they are smaller than the thickness of the line we use), although visible on very high resolution plots. For instance on the lower plot of Figure~\ref{fig:plni34}, the $n=2$ cusps appear at $q=\pm\sqrt{17}/4\  \mu\approx \pm 1.03\ \mu$. We have to stress that we positively verified also the appearance of all four new cusps for $E= 11/4\ \mu$ (they are at $\pm\sqrt{105}/4\ \mu$ and $\pm\sqrt{17}/4\ \mu$) but they are so faint, that presenting the appropriate drawings here is useless.

\subsection{$k=\pm 1$ instanton sectors}
\label{subsec:k1}

\begin{figure}[b]
\centering
{\includegraphics[width=0.45\textwidth]{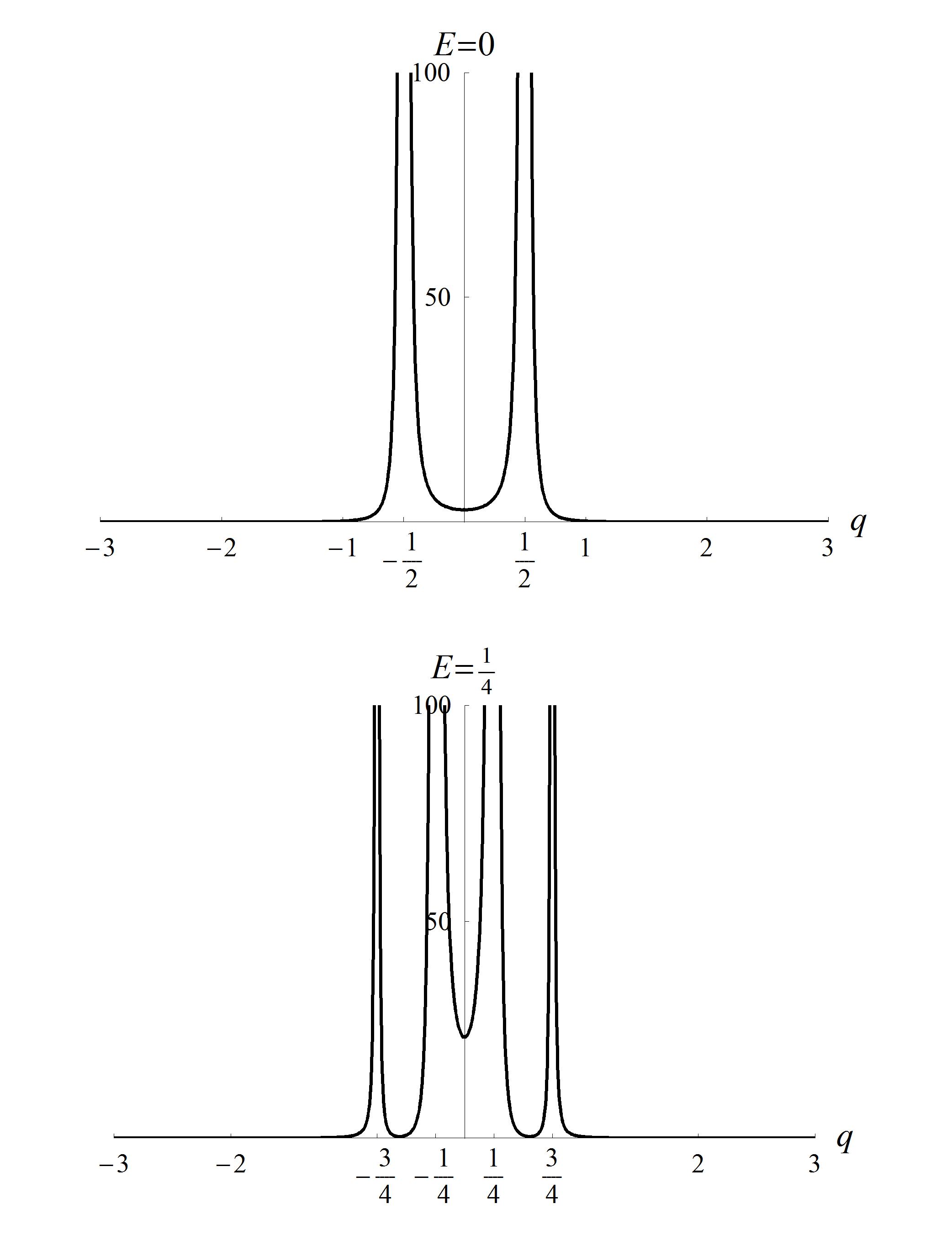}
\caption{Same as in Figure~\ref{fig:plni12} but for $|\Phi_P^{(1)}|$.} \label{fig:pli12}}
\end{figure}

To find out whether instanton terms exhibit the same threshold structure we have now to Fourier transform the expression~(\ref{eq:phip1}), which, however, similarly as for \mbox{$k=0$}, cannot be done explicitly. Fortunately an analogous integral representation for $Q^2<0$ as in~(\ref{eq:sspac}) may be used. If we define the quantity
\begin{equation}
B(p)=\int d^2 xe^{ipx}e^{ig^2\beta(x)},
\label{eq:bde}
\end{equation}
we obtain
\begin{eqnarray}
B(p)&\!\!\! =&\!\!\!-2\pi i\sqrt{\frac{2}{\mu}}\frac{1}{(-p^2)^{3/4}}\ e^{-\gamma_E/2}\label{eq:bspac}\\
&&\!\!\!\!\!\!\times\int_0^{\infty}dx \sqrt{x}J_0(x)\exp\left[-\frac{1}{2}K_0\left(\mu  
x/\sqrt{-p^2}\right)\right],
\nonumber
\end{eqnarray}
where $J_0$ is the Bessel function.

The instanton contribution to B-S function may be now written as
\begin{eqnarray}
\Phi_P^{(1)}(Q)&\!\!\! =&\!\!\!  \frac{\mu}{4\sqrt{\pi}}\ e^{\gamma_E}\gamma^5 e^{-i\theta\gamma^5}\label{eq:p1q}\\
&&\times\left[B\left(\frac{P}{2}+Q\right)+B\left(-\frac{P}{2}+Q\right)\right],
\nonumber
\end{eqnarray}

For the timelike $Q$, one finds
\begin{eqnarray}
B(p)&\!\!\! =&\!\!\!-\frac{2(1-i)}{\sqrt{\mu}}\frac{1}{(p^2)^{3/4}}\ e^{-\gamma_E/2}\label{eq:btim}\\
&&\!\!\!\!\!\!\times\int_0^{\infty}dx \sqrt{x}J_0(x)\exp\left[-\frac{i\pi}{4}H_0^{(1)}\left(\mu  
x/\sqrt{p^2}\right)\right].
\nonumber
\end{eqnarray}

\begin{figure}[h]
\centering
{\includegraphics[width=0.45\textwidth]{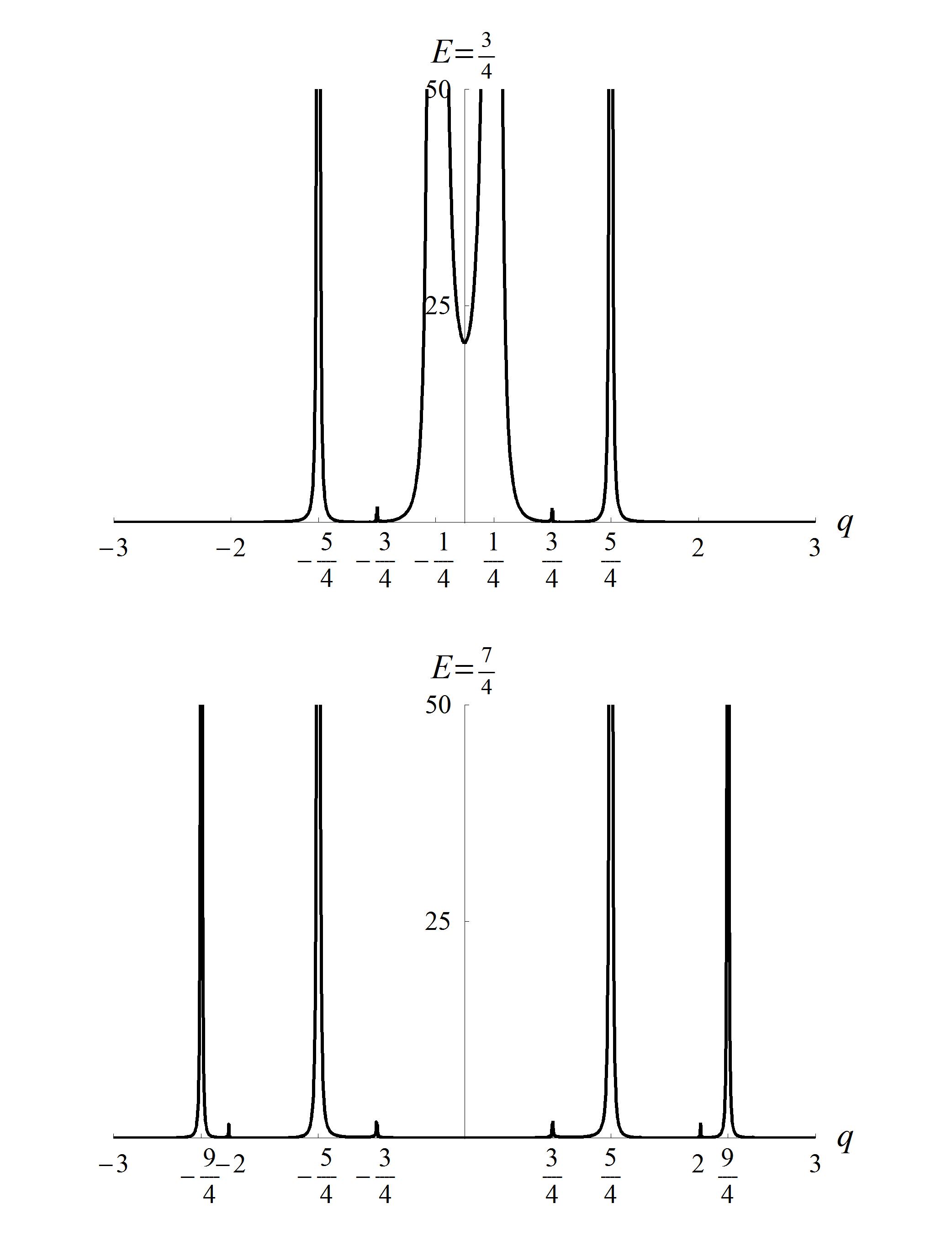}
\caption{Same as in Figure~\ref{fig:plni34} but for $|\Phi_P^{(1)}|$.} \label{fig:pli34}}
\end{figure}

\begin{figure*}
\centering
{\includegraphics[width=0.9
\textwidth]{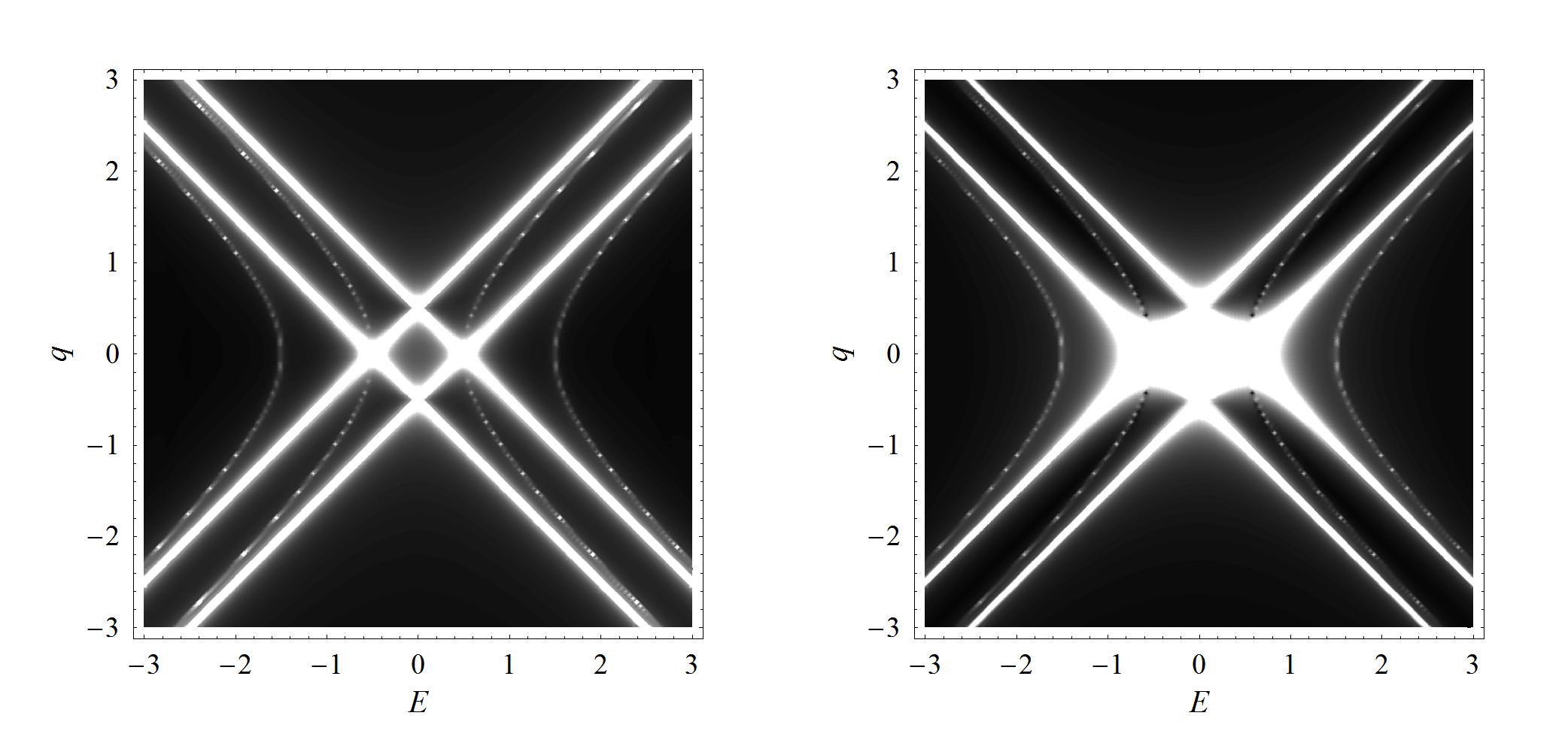}
\caption{The density graphics showing the behavior of $|\Phi_P^{(0)}|$ (left diagram) and $|\Phi_P^{(1)}|$ (right diagram). Bright areas correspond to regions of large values and dark areas of small ones. The colors are relative within each plot so they should not be compared between left and right diagram.} 
\label{fig:plic}}
\end{figure*}

As may be seen in Figures~\ref{fig:pli12} and~\ref{fig:pli34} the amplitude $|\Phi_P^{(1)}|$ has analogous kinematic singularities as $|\Phi_P^{(0)}|$, whose locations are dictated by~(\ref{eq:con0}). At the same threshold values of $E$ as before (i.e. $1/2\ \mu$, $3/2\ \mu$ for $n=1$ and $3/2\ \mu$, $5/2\ \mu$ for $n=2$ and so on) and for the same relative momenta the resonant cusps occur. Their positions are at $\pm 3/4\ \mu$ on the upper plot of Figure~\ref{fig:pli34} ($E=3/4\ \mu$) and at $\pm 3/4\ \mu,\ \pm \sqrt{65}/4\ \mu$ on the lower one ($E=7/4\ \mu$). They are now much less pronounced although still perfectly visible.

It is important to see, how these peaks move with relative two-momentum. If they really correspond to the Schwinger boson thresholds, they should be located on hyperbolas~(\ref{eq:ntre}) in the plane (E,q). That it is actually the case may be seen from the pictorial density diagrams of Figure~\ref{fig:plic} where light areas represent large values of the amplitudes, and dark small ones. Apart from the white lines, which show the locations of kinematic singularities, the two hyperbolas corresponding to $n=1$ thresholds are noticeable. We checked that the $n=2$ hyperbolas are also present, but their observation require very precise color plots and simultaneous artificial lowering the large values of $|\Phi_P^{(0,1)}|$ in order to reduce the maximal amplitude of colors.

The important question arises, how these effects emerge in~(\ref{eq:p0q}) and~(\ref{eq:p1q}) from the mathematical point of view. Why different expressions for $|\Phi_P^{(0)}|$ and $|\Phi_P^{(1)}|$ display almost identical threshold structure? The crucial observation is that infinite integrals~(\ref{eq:stim}) and~(\ref{eq:btim}) owe their   convergence to the presence of oscillating (phase) factors coming from Bessel and Hankel functions. If we expanded the exponential into a power series, the $n$-th term under the integral would contain a factor ($l$ stands for $0$ or $1$):
$$
J_l(x) \left[H_0^{(1)}\left(\frac{\mu}{\sqrt{p^2}}\ x\right)\right]^n.
$$
Now, in the large $x$ asymptotics, the Bessel function (both for $l=0$ and $l=1$) contributes $e^{\pm i x}$ to the phase factor, and $n$ Hankel functions $e^{i n \mu/\sqrt{p^2}\ x}$. If both factors cancel, i.e. if
\begin{equation}
n\ \frac{\mu}{\sqrt{p^2}}= \pm 1,
\label{eq:res}
\end{equation}
(of course only `$+$' comes into play) 
the value of integral over $x$ rapidly increases. The condition~(\ref{eq:res}) corresponds exactly to~(\ref{eq:ntre}). This again proves the correctness of our former observation, that the cusps are localized on hyperbolas in the relative two-momentum plane.

\section{Summary}
\label{sec:sum}

Summarizing the obtained results we would like to stress in the first place, that Schwinger Model turned out again to be exceptionally useful theory. The possibility of investigation {\em exact} bound-state wave-function in relativistic quantum field theory is unique even among model studies. Special mention deserves the fact that  it possesses two features typical to QCD: confinement and instantonic vacuum. 

The full Bethe-Salpeter wave-function in coordinate space in this model is known from the residue in the bound-state pole of the two-fermion Green's function, without the necessity of solving B-S equation itself (to tell the truth even the interaction kernel needed to formulate this equation is unknown). The contributions to the wave-function come from $k=0$ and $k=\pm 1$ instanton sectors (but not higher, since this quantity is defined through the operator bilinear in fermion fields). In section~\ref{sec:bsc} it was analyzed in the center-of-mass frame, how this wave-function behaves in relative position and in relative time. It turns out that it displays the light cone singularities and decreases as $1/\sqrt{r}$ for large space-like separations and $1/\sqrt{t}$ for time-like ones. This result may be, however, typical for $D=2$. As functions of relative time both noninstantonic and instantonic contributions show the oscillatory behavior shifted  by a phase of $\pi/2$ with respect to each other. Owing to that the total function defined by~(\ref{eq:topi}) is smoothed and has no nodes.

In the Fourier space we analyzed the behavior of $|\Phi_P^{(0)}|$ and $|\Phi_P^{(1)}|$ on relative momentum with fixed relative energy. In both instantonic sectors we found kinematic singularities coming from the external `legs' of the unamputated amplitude. For $E$ exceeding the threshold values there appear peaks corresponding to the Schwinger boson resonances. They emerge both for $k=0$ and $k=\pm 1$. They are located on hyperbolas defined by equations~(\ref{eq:ntre}). This threshold behavior has been suggested in approximated scalar-scalar model~\cite{keister}.

Since the Schwinger Model is on the one hand a nontrivial field theory and on the other is analytically solvable, the obtained results may be of certain importance for studies of more advanced QFT models. Both the retardation effects and going beyond `ladder' approximation are significant from the point of view of the correct description of the bound state~\cite{bha1}. The model opens also the possibility of investigations of the B-S amplitude in the boosted frame, the analytical properties in the complex relative energy plane and of form-factors.

\end{document}